\documentclass{article}

        \newcounter{eqnletter}[equation]
\catcode`\@=11 \@addtoreset{equation}{section} \catcode`\@=12

\usepackage[dvips]{graphicx}
\usepackage{color}
\usepackage{epsfig}

\begin{document}

  {\centerline	{\LARGE {  Sparse Random  Block Matrices : universality}} }
	 
	 \vskip .8cm
{\centerline {	Giovanni M. Cicuta$^*$
\footnote{giovanni.cicuta@gmail.com} and Mario Pernici$^{**}$
\footnote{mario.pernici@mi.infn.it} }} \vskip .3 cm {\centerline  { $^*$
Dept. of Physics, Univ. of Parma, Viale delle Scienze 7A, 43100 Parma,
Italy}} 
\vskip .3 cm
	{\centerline  {$^{**}$ Istituto Nazionale di Fisica Nucleare,
	Sezione di Milano ,}}
{	\centerline  {	Via Celoria 16, 20133 Milano, Italy}}

		\vskip 1cm {\centerline{\textbf{Abstract}} }
		\vskip .4 cm
 We study ensembles of sparse random block matrices generated from the
adjacency matrix of a Erd\"os-Renyi random graph  with $N$ vertices of average degree $Z$,
inserting a real symmetric $d \times d$ random block at each non-vanishing
entry. 

We consider some ensembles of random block matrices with rank $r < d$ and 
with maximal rank, $r=d$. 
The spectral moments of the sparse random block matrix are evaluated
for $N \to \infty$, $d$ finite or infinite, and several probability distributions for the blocks 
(e.g. fixed trace, bounded trace and Gaussian).
 Because of the concentration of the probability measure in the $d \to \infty$
limit, the spectral moments are independent of the probability measure of the blocks
(with mild assumptions of isotropy, smoothness and sub-gaussian tails).

The Effective Medium Approximation is the limiting spectral density
of the sparse random block ensembles with finite rank.
Analogous classes of universality hold for the Laplacian sparse block ensemble.
The same limiting distributions are obtained using random regular graphs
instead of Erd\"os-Renyi graphs.

	\vskip .6cm
	   \section{Introduction}
	 
The subject of this note is  an ensemble of sparse real
		symmetric block matrices, already studied in recent years.\\
		A generic block matrix of the ensemble is generated from a $N \times N$ matrix, see eq.(\ref{d.1}), after replacing each entry with a $d \times d$ random matrix $\alpha_{ij}X_{ij}$, $i,j=1,..,N$, as depicted in
		eq.(\ref{d.3}), then obtaining a $Nd \times Nd$ random block matrix.
		
				The set of random variables
				$\{\alpha_{ij}\}$ have distribution probabilities to make the random block
				matrix sparse, the set of blocks $\{X_{ij}
				\}$  are real symmetric random matrices
				of dimension $d \times d$. Our goal is
				the limiting, $ N \to \infty$, spectral
				distribution of the adjacency random block matrix $A$ in eq.(\ref{d.3}).
				The probability distribution of the random matrices  $\{X_{ij}
				\}$ and their relation to the universality properties of the spectral distribution of the ensemble $A$ are the central topic of this note.
				\\

Sparse random matrices are a difficult subject and many well known tools of random matrix theory cannot be used.
  Possibly the most
studied model of sparse random matrix ensemble is the adjacency matrix
of a non-oriented random graph with $N$ vertices of average vertex degree or average
connectivity $Z$.

\begin{eqnarray}
 A=\left( \begin{array}{cccccccc} 0 & \alpha_{1,2}
&\alpha_{1,3} & \dots &\alpha_{1,N} \\ \alpha_{2,1}& 0 &\alpha_{2,3}
& \dots & \alpha_{2,N} \\ \dots & \dots & \dots & \dots & \dots\\
\alpha_{N,1} &\alpha_{N,2} & \alpha_{N,3} &\dots & 0 \end{array}\right)
  \qquad , \qquad  \alpha_{i,j}=\alpha_{j,i} \qquad \qquad
\label{d.1} \end{eqnarray}

The set of $N(N-1)/2$ random variables $\{\alpha_{i,j} \}$
, $i>j$, is a set of independent identically distributed
random variables, each one having the probability density
\begin{eqnarray} P(\alpha)=\left(\frac{Z}{N}\right) \delta (\alpha
-1)+\left(1-\frac{Z}{N}\right) \delta(\alpha)\qquad \qquad \label{d.2}
\end{eqnarray}

The random matrix ensemble of eqs. (\ref{d.1}), (\ref{d.2}), was
considered a basic model of disordered system in statistical mechanics. It
was  analyzed for decades from the early days of the replica approach
\cite{pio} up to more recent cavity methods \cite{mod}, \cite{sla}, \cite{vivo}.\\

More pertinent to this paper, the moments of the spectral density  of
the limiting ($N \to \infty$) adjacency matrix $A$ in eq.(\ref{d.1}) were carefully studied
and recursion relations for them were obtained \cite{bau}, \cite{khor}.
Remarkably, the knowledge of all the spectral moments, at least in
principle, was not sufficient to obtain the spectral density.\\

In  recent years, an ensemble of sparse random block matrices was
considered, where the entry $A_{i,j}$ of the random matrix  is a real
symmetric $d \times d$ random matrix $X_{i,j}$ , \cite{pari}, \cite{cic1}, \cite{pern}, \cite{benet}, \cite{cic2},\\

\begin{eqnarray}
 A&=&\left( \begin{array}{cccccccc} 0 & \alpha_{1,2}
X_{1,2} &\alpha_{1,3} X_{1,3}& \dots &\alpha_{1,N} X_{1,N}\\
\alpha_{2,1}X_{2,1}& 0 &\alpha_{2,3} X_{2,3} & \dots & \alpha_{2,N}
X_{2,N}\\ \dots & \dots & \dots & \dots & \dots\\ \alpha_{N,1}X_{N,1}
&\alpha_{N,2} X_{N,2}& \alpha_{N,3}X_{N,3}&\dots & 0 \end{array}\right)
  \qquad , \nonumber
	 \end{eqnarray}
	\begin{eqnarray}
		X_{i,j}&=&X_{j,i}, \quad i,j=1, 2,..,N, \nonumber\\
	  X_{ij,\alpha \beta}&=& X_{ij, \beta\alpha }, \quad
	\alpha,\beta=1,2,..,d \qquad \qquad
\label{d.3} \end{eqnarray}

The generic block  $X_{i,j}$ may be considered a matrix weight associated
to the non-oriented edge $(i,j)$ of the graph. One may say that the set of
random variables $\alpha_{i,j} =\alpha_{j,i}$	encodes the architecture
of the non-oriented graph. In this case, see eq.(\ref{d.2}), it is the
Erd\"os-Renyi random graph, with average vertex degree (or connectivity)
$Z$.\\

It seems likely that the understanding of the spectral properties of sparse random block matrices of eq.(\ref{d.3}) has relevance on the dynamics of classes of networks. 
In several models of networks the set of nodes is partitioned into subsets, sometimes called communities. The interaction between any pair of nodes belonging to the same community is  different from the interaction of any pair of nodes belonging to different communities.	
	We recalled in Appendix B of \cite{cic2} some similarities of the stochastic block model and the Equitable random graph with the sparse block matrix ensembles here studied. References quoted there may introduce readers to the vast literature of complex systems modeled on random networks.\\
	
Our goal is the evaluation of the limiting moments $\mu_p$ of the sparse random block matrix $A$ in  eq.(\ref{d.3}), (Tr is the trace in the $Nd$-dimensional space, tr is the trace in the $d$- dimensional space).
	
		 \begin{eqnarray}
\mu_{p}=\lim_{N \to
\infty} \frac{1}{Nd}<\texttt{Tr} A^{p}> \qquad \qquad
\label{d.4} \end{eqnarray}

	Moments $\mu^p$ are evaluated in terms of weighted paths with $p$ steps on a complete graph.
		In the present case the
	weight of any path is the trace of the product of the matrices
	associated to the edges. Because of the probability distribution of the $\{ \alpha_{i,j} \}$ random
variables, eq.(\ref{d.2}), in the limit $N \to \infty$, only closed paths
on trees contribute to the limiting moments, \cite{bau}, \cite{khor}.\\
This holds true regardless the probability distribution of the random
blocks $X_{i,j}$ and their finite dimension $d$.\\ For a closed  path on
a tree,  every edge is traversed an even number of times. The limiting
moments of odd order vanish.\\ The weight of a path is the product of
the weights of the traversed edges. For a closed walk of $2p$ steps,
the number $l$ of distinct matrices occurring in the weighted path is $1
\leq l \leq p$.\\
 Since the blocks are i.i.d., the identification of a
block $X_{i,j}$ in a product is irrelevant.  It is useful a relabeling
of the products of the blocks that only records if the blocks are equal
or different to other  ones in the product. For instance :

\begin{eqnarray} &&X_{1,3}X_{3,1}X_{1,3}X_{3,4}
X_{4,7}X_{7,4}X_{4,3}X_{3,1}\qquad \qquad \texttt{is relabeled} \qquad
\nonumber\\ && (X_1)^3 X_2 (X_3)^2  X_2 X_1 \qquad \qquad \label{d.6}
\end{eqnarray}

As the order of the spectral moments increases, the number of relevant
products also increases. 
In Appendix A of \cite{cic2}, we listed the analytic contributions
	 up to $\mu_{10}$. It holds for any probability
	choice for the random matrices $X_{i,j}$ and any dimension $d$, 
	provided that the
	random block matrix ensemble $A$ has the Erd\"os-Renyi  structure
	and the matrix blocks are i.i.d.\\
	 
 In the next Section we consider ensembles of random block matrices of fixed rank $r$,
parametrized by $r$ random vectors, in which the vectors have uniform distribution
on a sphere, uniform distribution in a ball, or a gaussian measure;
then we consider random block matrices of maximal rank, with fixed trace, bounded trace
and Gaussian distribution.

We  show that the $d \to \infty$ limit with $\frac{Z}{d}$ limit of a
fixed rank model is the same regardless  the  class of probability distributions
considered for the vectors, and that the $d \to \infty$ limit of the maximal rank model 
is the same regardless  the considered class of probability distributions for the blocks.
The proof is analogous in all these cases. This universality can be traced back
to known properties of high-dimension probability; we recall the general conditions
which lead  to the concentration of probability measures and the universality properties
of the spectral moments.

In the case ot the finite rank models
the spectral distribution is the Effective Medium distribution with parameter
$t = \frac{r Z}{d}$.

\section{Expectations}
As it is mentioned in the Introduction, the spectral moments of eq.(\ref{d.4}) are polynomials in the $Z$ variable, with increasing number of different blocks. For instance
\begin{eqnarray}
\frac{1}{N}  \texttt{Tr}
\,A^8 &=& Z\,\texttt{tr}\,X_1^8+Z^2\,\texttt{tr}\,\left[ 8\,X_1^6X_2^2
  +4\,X_1^4X_2^4+2\,X_1^2X_2^2X_1^2X_2^2\right]+\nonumber\\
  && Z^3\,\texttt{tr}\,\left[8\,X_1^4X_2^2X_3^2+8\,X_1^4X_2X_3^2
  X_2+8\,X_1^3X_2^2 X_1X_3^2+
	4\,X_1^2X_2^2X_1^2X_3^2\right]+\nonumber\\
	&& Z^4\,\texttt{tr}\, \left[8\,X_1^2X_2^2 X_3X_4^2 X_3+
	4\, X_1^2 X_2X_3X_4^2 X_3X_2+2\, X_1^2X_2^2
	X_3^2X_4^2 \right] \nonumber\\
	\label{e.1}
\end{eqnarray}
The highest power of $Z$ in the polynomial multiplies the contribution of the Wigner paths, where each traveled edge is traveled exactly twice.\\
The expectations of each term will be shown to be independent of the probability measure generically (for subgaussian measures), in the $d \to \infty$ limit.

	\subsection{rank one random blocks $X$}
		
		We begin considering an ensemble of $d \times d$ real symmetric random matrices $X_{i,j}$  independent
  (except for the symmetry $X_{i,j}=X_{j,i}$) identically distributed. Each matrix is function of just a $d$-dimensional random vector
		 $ \vec{v} \in
  R^d$ ,  $X=|\vec v><\vec v|$. \\ We consider three  probability measures for the random vector.\\
	
	-Uniform probability on the sphere
	\begin{eqnarray}
		P_{\delta} (\vec v)= \frac{1}{Z_\delta} \delta \left( R^2-\sum_{j=1}^d v_j^2\right)  \qquad , \qquad
		Z_\delta =  \frac{1}{2R} S_d(R)=\frac{  \pi^{d/2}}{\Gamma \left(\frac{d}{2}\right)} R^{d-2}
		\qquad \qquad \label{p.1}
 \end{eqnarray}
		Than the random matrix $X$ is a projector into a one-dimensional space. This probability
  distribution for the set $\{X_{i,j} \}$ is  unusual in random matrix
  literature  \footnote{ It is motivated by  physics : the random Laplacian block
  matrix associated to the random block matrix $A$ in eq.(\ref{d.3}) is
  the Hessian of a system of points of random locations,  connected by
  springs, \cite{pari}, \cite{cic1}, \cite{pern}, \cite{benet}.}.
	In the limit $d \to \infty$, with $t=Z/d$ fixed, moments of all orders are evaluated \cite{pern}, the limiting resolvent is
solution of a cubic equation, previously obtained in a different model and
different approximation, the Effective Medium Approximation \cite{semer}.\\

-Uniform probability on the ball
	\begin{eqnarray}
		P_{\Theta} (\vec v)&=& \frac{1}{Z_\Theta} \Theta \left( R^2-\sum_{j=1}^d v_j^2\right)  \qquad , \nonumber\\
		Z_\Theta &=& \int_{\sqrt{\sum_1^d v_j^2} \leq R } \, \prod_{j=1}^d  dv_j=V_d(R)=\frac{  \pi^{d/2}}{\Gamma \left(\frac{d}{2}+1\right)} R^{d}
		\qquad \qquad \label{p.2}
 \end{eqnarray}

	-Gaussian measure
	\begin{eqnarray}
		P_{\textsl{gauss}}(\vec v) &=& \left(\frac{d}{2 \pi R^2}\right)^{d/2}  e^{-\left(\frac{d}{2R^2}\right) (\vec v \cdot \vec v)}   
		\qquad \qquad \label{p.3}
 \end{eqnarray}

Let us consider the trace of the product of $P$ random matrices  $X$, where $r_1=m_1+m_2+\dots$ is the sum of the powers of the matrix $X_1$ in the product,
$r_2=n_1+n_2+\dots$ is the sum of the powers of the matrix $X_2$ in the product, etc., $s$ is the number of distinct matrices in the product,
$P=\sum_{k=1}^s r_k$, each $r_k$ is an even integer.

$$\texttt{tr}\, X_1^{m_1}X_2^{n_1}\dots X_1^{m_2}\dots X_3^{p_1} \dots$$

Expectations of traces of products of the random matrices $\{X\}$ with the three probability measures eq.(\ref{p.1})-(\ref{p.3}) are straightforward. We merely quote here two equations which compare expectations of the uniform probability on the ball with the Gaussian measure and
the uniform probability on the sphere  with the Gaussian measure. They are valid for any $R$ and $d$ and  may be obtained  by Laplace transform, as it is described in the third subsection. The method and the results are a simple multi-matrix generalization of well known fact, beginning with Rosenzweig and Bronk \cite{rosen}, later elaborated and generalized \cite{gern}, \cite{dela}, \cite{gg}, \cite{kopp}.\\
 \begin{eqnarray}
< \texttt{tr}\Bigg( X_1^{m_1} X_2^{n_1}..X_1^{m_2}..X_3^{p_1}..\Bigg)>_{\Theta}&=& F_{\Theta}( d) 
< \texttt{tr}\Bigg( X_1^{m_1} X_2^{n_1}..X_1^{m_2}..X_3^{p_1}..\Bigg)>_\textsl{gauss}\qquad ,  \nonumber\\
 F_{\Theta}( d) &=& \prod_{k=1}^s \frac{d^{r_k}\Gamma\left( \frac{d}{2}+1\right)}{2^{r_k} \Gamma \left( \frac{d}{2}+r_k+1\right)}
\quad \label{t.1}
\end{eqnarray}

\begin{eqnarray}
< \texttt{tr}\Bigg( X_1^{m_1} X_2^{n_1}..X_1^{m_2}..X_3^{p_1}..\Bigg)>_{\delta}&=& F_{\delta}( d) 
< \texttt{tr}\Bigg( X_1^{m_1} X_2^{n_1}..X_1^{m_2}..X_3^{p_1}..\Bigg)>_\textsl{gauss}\qquad  ,   \nonumber\\
 F_{\delta}( d) &=& \prod_{k=1}^s \frac{d^{r_k}\Gamma\left( \frac{d}{2}\right)}{2^{r_k} \Gamma \left( \frac{d}{2}+r_k\right)}
\quad \label{t.2}
\end{eqnarray}

 Since
$$\lim_{d \to \infty} F_{\Theta}(d)=1 \quad , \quad \lim_{d \to \infty} F_{\delta}(d)=1 $$
the three probability measures, $P_\delta$, $P_\Theta$, $P_{\textsl{gauss}}$, obtain the same expectations for any multi-matrix product in the $d \to \infty$ limit. The fact that  the \textsl{volume} probability distribution $P_{\Theta}$ obtains the same expectations of the \textsl{surface} probability distribution $P_{\delta}$, is the most simple  example of concentration of the probability measure in spaces of high dimension. Any probability measure for the components of the random vector, such that the probability on the tails is bound by a Gaussian function (the sub-gaussian distributions) would also lead to the concentration of the measure in the $d \to \infty$ limit \cite{ledoux}, \cite{versh}.\\

With the probability $P_\delta$, it was proved in \cite{cic1} for moments of low order and in \cite{pern} for moments of every order that in the limit $d\to \infty$ and $Z \to \infty$ with fixed ratio $t=Z/d$ the products of multi-matrices associated with non-crossing partitions have finite limit whereas the products of multi-matrices associated with crossing partitions vanish. This allows the determination of the non random spectral distribution of the matrix $A$ in this limit, the resolvent obeys a cubic equation, sometimes called Effective Medium Approximation, quoted in eq.(\ref{em.1}).
In the same limit, it was shown that the spectral density of the associated random block Laplacian is the Marchenko-Pastur density, quoted in eq.(\ref{em.2}).
Our work was confirmed by an independent derivation
 \cite{aa}. We may now assert, by the measure concentration, that infinitely many probability distributions of the random blocks $\{X\}$ lead, in this limit, to the same spectral distribution for the sparse random block matrix $A$, the crucial feature being the blocks having 
finite rank  , as it is shown in next subsection. \\ 

\subsection{ random blocks $X$ of  rank $r$}

The derivations related to the random blocks $X$ of unit rank, may be generalized to random blocks $X$ of any rank, provided it remains finite, in the limit $d \to \infty$, $Z \to \infty$, $t=Z/d$ fixed.\\
	We  consider an ensemble of $d \times d$ real symmetric random matrices $X_{i,j}$  independent
  (except for the symmetry $X_{i,j}=X_{j,i}$) identically distributed. Each matrix is function of just a set of $d$-dimensional random vectors
		 $ \{\vec v^{(a)} \}\in
  R^d$ ,  $X=\sum_{a=1}^r|\vec v^{(a)}><\vec v^{(a)}|$. \\ 
	The vectors are orthogonal, $(\vec v^{(a)}\cdot\vec v^{(b)})=0$ if $ b \neq a$. The random matrix $X$ has  rank $r<d$, it is function of $r(2d-r+1)/2$ independent coordinates and
		$$X^m=\sum_{a=1}^r (\vec v^{(a)}\cdot\vec v^{(a)})^{m-1}|\vec v^{(a)}><\vec v^{(a)}|\quad , \quad \texttt{tr} \,X^m= \sum_{a=1}^r (\vec v^{(a)}\cdot\vec v^{(a)})^{m}.  $$\\
	
	We consider the    joint probability measure for the random vectors
	\begin{eqnarray}
	P(\vec v^{(1)},..,\vec v^{(r)})&=&\frac{1}{Z}\prod_{a=1}^r \delta \left(R^2-(\vec v^{(a)} \cdot \vec v^{(a)})\right) \prod_{i<j} \delta(\vec v^{(i)} \cdot \vec v^{(j)}) \nonumber\\
	Z&=& \left( \prod_{j=1}^r \frac{ \Omega_{d-j}}{2}\right) \, R^{r(d-r-1)} \, , \, \Omega_{d-1}=2 \frac{\pi^{d/2}}{\Gamma(d/2)} \qquad
	\label{r.1}
 \end{eqnarray}
	
	This probability measure will exhibit the consequences of the finite rank $r$ and the large dimensionality $d \to \infty$. Integration over sets of orthogonal vectors is mentioned in chap. 21 of \cite{meh} and the paper \cite{ym}.\\
	
	In the limit $d \to \infty$ the expectations are evaluated in a way completely analogous to the case of the probability density in eq.(\ref{p.1}) and in the paper \cite{pern}, by distinguishing contributions to the moments associated to non-crossing partitions from those associated to the crossing partitions. For instance, all the terms, except one, in eq.(\ref{e.1}) are associated to non-crossing partitions. They are evaluated by repeated use of factorization  \footnote{ In the paper \cite{cic2}, where the blocks $\{X_j\}$ are member of the Gaussian Orthogonal Ensemble, we recalled the relevance of factorization to evaluate expectations of multi-matrix products for $d$ finite or infinite.
	If the entries of the pair of matrices $A$ , $B$, are stochastically independent
	 $$ <(A B)_{r,s}>=\sum_{j=1}^d <A_{r,j}><B_{j,s}>$$   Next the isotropy relation
	$$< (X_j^k)_{r,s}>=\delta_{r,s} \frac{1}{d} <\texttt{tr} X_j^k> $$  leads to
	$$<\texttt{tr}\,X_1^6X_2^2>= \frac{1}{d} <\texttt{tr}\,X_1^6  >< \texttt{tr}\,X_2^2>$$
	}. For instance
 	\begin{eqnarray}
	Z^4\, \frac{1}{d}\texttt{tr}  < X_1^2 X_2 X_3 X_4^2 X_3 X_2>&=&Z^4 \Bigg( \frac{1}{d}<\texttt{tr} X_4^2> \frac{1}{d}\texttt{tr}<X_1^2 X_2 X_3^2 X_2>\Bigg) \nonumber\\
	&=& Z^4  \left( \frac{1}{d}\texttt{tr}  < X^2> \right)^4=\left(\frac{Z r}{d}\right)^4 R^{16}
	\nonumber\\
	Z^2\,\frac{1}{d} <\texttt{tr}\,\left[ 8\,X_1^6X_2^2
  +4\,X_1^4X_2^4 \right] >&=& 12\left(\frac{Z r}{d}\right)^2 R^{16}
	\nonumber
	\end{eqnarray}

	The only term  in eq.(\ref{e.1}) associated to crossing partitions may be neglected, in the $d \to \infty$ limit because it involves higher number of internal products $<\vec v|\vec w>$ between distinct vectors
	\begin{eqnarray}
\frac{1}{d}  
  Z^2\, <\texttt{tr}\,  \,X_1^2X_2^2X_1^2X_2^2 > \sim  O\left(\frac{Z^2}{d^3}\right) R^{16}
   \nonumber
\end{eqnarray}
	
	 All the non-crossing contributions may be evaluated, as in the case of rank one blocks. The spectral moments of the sparse random blocks matrix $A$, in the limit $d \to \infty$, with the ratio $Z/d$ fixed, reproduce the moments of the Effective Medium Approximation, with the parameter $t=rZ/d$.\\

We describe now a similar ensemble of random blocks $X$ each one made of $r$  independent random vectors $\vec v^{a} \in R^d$, without the constraint of being an orthogonal set.
Since for $d \to \infty$ any two random vectors tend to be orthogonal, one expects that this ensemble, for $d \to \infty$, reproduces the result of the ensemble of orthogonal vectors and it is easier for simulations.\\
 We now define the $d \times d$ blocks $X$ of rank $r$ and a factorized joint probability distribution

	\begin{eqnarray}
X&=&\sum_{a=1}^r  |\vec v^{(a)}><\vec v^{(a)}| \qquad ,\nonumber\\
	P(\vec v^{(1)},..,\vec v^{(r)})&=&\frac{1}{Z}\prod_{a=1}^r \delta \left(R^2-(\vec v^{(a)} \cdot \vec v^{(a)})\right)  \quad , \qquad
		Z =  \left(\frac{  \pi^{d/2}}{\Gamma \left(\frac{d}{2}\right)} R^{d-2}\right)^r \nonumber\\
	 	\label{u.1}
 \end{eqnarray}

We begin by examining the expectation of the trace of the power of a single block in the limit $d \to \infty$ 
\begin{eqnarray}
\lim_{d \to \infty} \frac{1}{d} <\texttt{tr} \, \alpha X^m>=\frac{Z}{d} <\left( \sum_{a=1}^r (\vec v^{(a)} \cdot \vec v^{(a)})^m+ \sum \prod (\vec v^{(b)} \cdot \vec v^{(c)}) \right) \qquad
\label{u.2}
\end{eqnarray}
The term $\sum \prod$ represents products of pairs of different vectors times products of pairs of the same vectors. We recall ( see for instance eqs.(9)-(11) in ref.\cite{pern} ) that expectations of scalar products vanish in the limit $d \to \infty$. Then
\begin{eqnarray}
\lim_{d \to \infty} \frac{1}{d} <\texttt{tr} \, \alpha X^m>=\frac{Z}{d} r R^{2m}
\label{u.2b}
\end{eqnarray}

A mild generalization of the second part of Proposition 1 in ref.\cite{pern} allows to prove that the expectation of the trace of a product of any number of blocks, which correspond to a non-crossing partition, gives the contribution $t^m$ , where $m$ is the number of distinct blocks. \\
Indeed in a non-crossing term, there  exist at least a block $X_j$ appearing as $(X_j)^k$ in one position and not elsewhere. Averaging over the vectors of the block $X_j$ we find, as in eq.(\ref{u.2})
\begin{eqnarray}
<\vec v |\alpha(X_j)^k| \vec w>&=&<\vec v |\alpha R^{2(k-1)} X_j +\sum \prod (\vec v^{(b)} \cdot \vec v^{(c)})   | \vec w> \nonumber\\
& \sim & r Z R^{2 k} (\vec v \cdot \vec w)+ O(1/d)
	\label{u.3}
 \end{eqnarray}
The resulting product of blocks is still non-crossing, the steps may be repeated with a new block $X_i$, until one is left with a single block, like in eq.(\ref{u.2b}).\\

Finally it remains to show that the average of the trace of a product of blocks, which is crossing, is negligible in the $d \to \infty$ limit.\\
A generic product of blocks, which is crossing, may  contain one or more blocks $X_j$ appearing as powers $(X_j)^k$ in one position of the product and nowhere else. By repeating the steps of eq.(\ref{u.3}) one eventually obtains the expectation of a \textsl{reduced} product of blocks, still crossing, where no block occurs in a single position. Its asymptotic behaviour in the $d \to \infty$ limit, is the same of the original product.\\
Every block $(X_i)^{m_i}$ in the \textsl{reduced} product occurs at least in two non consecutive positions and its random vectors form scalar products with a larger number of distinct vectors than it would happen in case of consecutive positions. Then all crossing contributions may be neglected in the limit $ d \to \infty$, $Z \to \infty$, with fixed ratio $Z/d$.\\
A detailed evaluation of the contributions is done in the first part of Proposition 1 in ref.\cite{pern}.\\

Remarks.\\
The spectral moments $\mu_{2k}$ of the sparse random block matrix, with $R=1$, in the limit $ d \to \infty$, $Z \to \infty$, with fixed ratio $Z/d$, are those of the Effective Medium Approximation. The generating function of the moments, $g(z)=\sum_{k=0} \frac{\mu_{2k} }{z^{2k+1}}$,  is solution of the cubic equation
\begin{eqnarray}
 [g(z)]^3+\frac{t-1}{z}[g(z)]^2-g(z)+\frac{1}{z}=0 \qquad , \qquad t= \frac{r\,Z}{d} \qquad
\label{em.1}
 \end{eqnarray}

The spectral moments  of the sparse random block Laplacian matrix, in the same limit are the spectral moments of the Marchenko-Pastur distribution
$\rho_{MP}(x)$
\begin{eqnarray}
\rho_{MP}(x)&=&\frac{ \sqrt{(b-x)(x-a)}}{4\pi \, x} \qquad , \qquad 0 \leq a \leq x \leq b \qquad \nonumber\\
a&=&(\sqrt{t}-\sqrt{2} )^2 \quad , \quad b=(\sqrt{t}+\sqrt{2} )^2 \qquad , \qquad t= \frac{r\,Z}{d}
\qquad \qquad
\label{em.2}
\end{eqnarray}

 We performed a few simulations with random block matrices with blocks with
dimension $d$ between $2$ and $8$,
both for the adjacency and the Laplacian sparse block ensembles,
with blocks of rank $r=1$ with the measure with the gaussian probability,
and with blocks of rank $r=2$, with independent random vectors with
uniform probability on the sphere.
They support the analysis and the conclusions of this note. 
The spectral distribution approaches the limiting distribution as
$d$ increases, but slower than in the case $r=1$ with uniform probability
on the sphere and same $t$.

For sake of simplicity we considered in eqs.(\ref{r.1}) and (\ref{u.1}) only the fixed length probability of the random vectors. Also the bounded length or the gaussian length may be considered. The blocks $\{X \}$ are sum of $r$ contributions, each one being of the form considered in rank one blocks subsection. The equivalence of the different measures in the $d \to \infty$ limit is established by the method discussed there.\\

The  fixed length probability of the random vectors considered in eq. (\ref{u.1}) may be generalized by associating different $R_a$ to the vectors $\vec v^{a)}$ and the limiting resolvent of the Adjacency block ensemble satisfies a polynomial equation of order higher  than $3$. The use of 
the non-crossing partition transform, described in sect.4 of \cite{pern} is convenient: only paths associated to non-crossing partitions are relevant, the generating function is
\begin{eqnarray}
a(x)=
\sum_{k \geq 1}<\texttt{tr} \, \alpha X^{2k}>x^{2k}=\sum_{k \geq 1} \frac{Z}{d}\sum_{a=1}^r R_a^{4k}x^{2k}= \frac{Z}{d}\sum_{a=1}^r \frac{x^2 R_a^4}{1-x^2 R_a^4}
\label{rem.1}
 \end{eqnarray}
Using the non-crossing  partition transform, the generating functions of the moments is
\begin{eqnarray}
f(x)=1+a\left(xf(x)\right)=1+\frac{Z}{d}\sum_{a=1}^r \frac{x^2 f^2(x) R_a^4}{1-x^2 f^2(x)R_a^4}
\label{rem.2}
 \end{eqnarray}
We did not attempt to compute the spectral distribution in this more general case.

\subsection{ random blocks $X$ of maximum rank}

We consider ensembles of random real symmetric matrices $\{X\}$ of order $d$, with three probability measures  analogous to the ones in eqs.(\ref{p.1})-(\ref{p.3})\\

	- Fixed trace
	\begin{eqnarray}
		P_{\delta} (X)&=& \frac{1}{Z_\delta} \delta \left( R^2-\frac{1}{d} \texttt{tr}X^2 \right)  \qquad , \nonumber\\
		Z_\delta &=&  \frac{  (\pi d)^{d(d+1)/4}}{2^{d(d-1)/4}\Gamma \left(\frac{d(d+1)}{4}\right)} R^{d(d+1)/2-2}
		\qquad \qquad \label{b.1}
 \end{eqnarray}

- Bounded trace
	\begin{eqnarray}
		P_{\Theta} (X)&=& \frac{1}{Z_\Theta} \Theta \left( R^2-\frac{1}{d} \texttt{tr}X^2 \right)  \qquad , \nonumber\\
		Z_\Theta &=& \int_{ \frac{1}{d}\texttt{tr}X^2 \leq R^2 } \, \prod_{i \leq j}^d  dX_{i,j}=\frac{  (\pi d)^{d(d+1)/4}}{2^{d(d-1)/4}\Gamma \left(\frac{d(d+1)}{4}+1\right)} R^{d(d+1)/2}
		\qquad \qquad \label{b.2}
 \end{eqnarray}

- Gaussian measure
\begin{eqnarray}
		P_{\textsl{gauss}} (X)= \frac{1}{Z_{ \textsl{gauss}}}  e^{-(d/4 R^2)  \texttt{tr}X^2}   \qquad , \qquad
		Z_\textsl{gauss} =  2^{d/2} \left(\frac{ R \sqrt{2\pi}}{\sqrt{d} }\right)^{d(d+1)/2}
		 \qquad  \qquad 
		\label{b.3}
 \end{eqnarray}

We quote here two equations which compare expectations of the trace of a generic multi-matrix product for the fixed trace probability in eq.(\ref{b.1})  with the Gaussian measure in eq.(\ref{b.3}) and for
the bounded trace probability in eq.(\ref{b.2})  with the Gaussian measure (\ref{b.3}). They are valid for any $R$ and $d$.\\

 \begin{eqnarray}
< \texttt{tr}\Bigg( X_1^{m_1} X_2^{n_1}..X_1^{m_2}..X_3^{p_1}..\Bigg)>_{\Theta}&=& F_{\Theta}( d)
< \texttt{tr}\Bigg( X_1^{m_1} X_2^{n_1}..X_1^{m_2}..X_3^{p_1}..\Bigg)>_\textsl{gauss}  \quad , \quad \nonumber\\
 F_{\Theta}( d) &=& \prod_{k=1}^s \frac{d^{r_k}\Gamma\left( \frac{d(d+1)}{4}+1\right)}{4^{r_k/2} \Gamma \left( \frac{d(d+1)}{4}+\frac{r_k}{2}+1\right)}
\qquad \label{b.5}
\end{eqnarray}

\begin{eqnarray}
< \texttt{tr}\Bigg( X_1^{m_1} X_2^{n_1}..X_1^{m_2}..X_3^{p_1}..\Bigg)>_{\delta}&=&F_{\delta}( d) 
< \texttt{tr}\Bigg( X_1^{m_1} X_2^{n_1}..X_1^{m_2}..X_3^{p_1}..\Bigg)>_\textsl{gauss}\quad , \quad \nonumber\\
 F_{\delta}( d) &=& \prod_{k=1}^s \frac{d^{r_k}\Gamma\left( \frac{d(d+1)}{4}\right)}{4^{r_k/2} \Gamma \left( \frac{d(d+1)}{4}+\frac{r_k}{2}\right)}
\qquad \label{b.6}
\end{eqnarray}

An outline of the derivation of eqs.(\ref{b.5}), (\ref{b.6}) is the following. One inserts the integral representation
 \begin{eqnarray}\Theta \left( R_j^2-\frac{1}{d} \texttt{tr}X_j^2 \right) = \frac{1}{2\pi i} \int_{-\infty}^{\infty} \frac{dy_j}{y_j-i\epsilon} e^{iy_j \left( R_j^2-\frac{1}{d} \texttt{tr}X_j^2 \right)}
 = \frac{1}{2\pi i} \int_{\epsilon-i\infty}^{\epsilon+i\infty} \frac{dz_j}{z_j} e^{(z_j-\epsilon)\left( R_j^2-\frac{1}{d} \texttt{tr}X_j^2 \right)} 
\nonumber
\end{eqnarray}

for each of the distinct blocks, $j=1,2,..,s$, occurring in the integral
$$ \int \texttt{tr}\Bigg( X_1^{m_1} X_2^{n_1}..X_1^{m_2}..X_3^{p_1}..\Bigg) \prod_{j=1}^s \Theta \left( R_j^2-\frac{1}{d} \texttt{tr}X_j^2 \right) dX_j$$

One performs the gaussian multi-matrix integral over the distinct blocks $X_j$, next the integration of the $z_j$ variables in the complex plane; finally  the division of the $s$ normalization factors $Z_{\Theta}$ lead to eq. (\ref{b.5}). In the same way one obtains eqs.(\ref{t.1}), (\ref{t.2}), (\ref{b.6}). They are straightforward  multi-matrix generalizations of one matrix equations known long ago \cite{rosen}, \cite{gern}, \cite{dela}, \cite{gg}, \cite{kopp}. Analogous equations hold for complex hermitian blocks $X_{i,j}$.\\
 Since
$\lim_{d \to \infty} F_{\Theta}(d)=1  , \quad \lim_{d \to \infty} F_{\delta}(d)=1 $, 
the three probability measures, $P_{\delta}$, $P_{\Theta}$, $P_{\textsl{gauss}}$, eqs.(\ref{b.1})-(\ref{b.3}), lead to
 the same expectations for any multi-matrix product in the $d \to \infty$ limit. At the time of the derivation of equal results for any moment, for the one matrix case, in the $d \to \infty$ limit, it did seem related to the special analytic relations between Gaussian probability and restricted trace probabilities. Now it is understood as an example of concentration of the probability measure.\\

\section{Classes of universality and other ensembles}

In this note we call universality the property that the spectral moments of the  sparse random block matrix $A$ converge to the same deterministic limit,
in the limit: $\lim_{d \to \infty} \lim_ {N \to \infty}$, regardless
 the probability measure of the entries of the blocks $X_{i,j}$ .\\
Concentration inequalities proved in spaces of high dimensionality lead to these results. Indeed the set of real random independent entries of the random matrix $X$ may be seen as a vector in a real space with $d(d+1)/2$ dimensions. If we assume that each component of this vector has a probability density sub-gaussian, mean zero and unit variance, then the joint distribution of the entries, in the $d \to \infty$ limit concentrates close to the thin spherical surface of radius $\sqrt{d(d+1)/2}$, see  \cite{ledoux}, and ch.3 \textsl{Random vectors in high dimensions} in \cite{versh}.\\
Furthermore the same concentration inequalities allow to evaluate any smooth (Lipschitz) functional of the random variables on the concentrated measure. The expectation of each trace of product of blocks and their sum,
all the spectral moments of the sparse block matrix $A$ obtain limiting values independent of the probability distribution and the limiting spectral density of the matrix itself has this universality property. If the blocks $X$ are random matrix with full rank, we do not know the limiting spectral density. 
If the blocks
 $X$ have finite rank $r \geq 1$, the set of walks on trees on the random graph which contribute to the moments is the set associated to non-crossing partitions. In the most simple cases, with the spherical symmetry considered in this note, the limiting universal density is the Effective Medium Approximation with a parameter $t=rZ/d$.\\

 The limiting deterministic spectral density may depend on the symmetries of the joint probability distribution of the entries of the random matrices. A simple example is provided by the rank-one blocks $X=|\vec x><\vec x|$ where the random $d$-vector $\vec x$ has uniform distribution in the cube $[-R,R]^d$. This probability measure concentrates to the "`skin"' of the cube, that is its surface.\\
It seems proper to consider classes of universality, since infinite probability distributions of the matrix entries, with the same rank of the random matrix, the same symmetry properties and sub-gaussian tails of the distributions are expected to concentrate to the same joint probability distribution in the $d \to \infty$ limit, then obtaining the same limiting expectations.\\

Analogous results are obtained for Laplacian sparse random blocks. Universality of the spectral moments is assured by the sub-gaussian distribution of the entries of the blocks $X$. The limiting spectral function, the Marchenko-Pastur distribution, is obtained if the blocks
 have any finite rank $r \geq 1$,  see \cite{cic1}-\cite{aa} for the case $r=1$ with
random vectors with uniform distribution on the sphere.\\  
In Appendix A of \cite{cic2}, it was mentioned how to modify the multiplicities of the products of blocks corresponding to walks on trees, in order to obtain the moments of regular random block matrices. Here too, universality of the expectations implies that the limiting spectral density of the random regular block ensemble is not dependent on the probability distribution of the block entries.  
In the case of blocks of finite rank $r \geq 1$, the limit $d \to \infty$ is performed with the degree $Z \to \infty$ and fixed ratio $t=r Z/d$.
 The limiting spectral function is then the same Effective Medium Approximation obtained for the sparse block random ensemble $A$. If the blocks have maximum rank, the limiting spectral density of the random regular block ensemble is universal, but not known.\\

\section{Conclusions}

By some explicit analytic evaluations and some implications of the theory of probability in high dimensional spaces, the spectral moments of ensembles of sparse random block matrices are shown to be independent of the probability measure of the blocks in the double limit $N \to \infty$ first and $d \to \infty$ next. The evaluations crucially depend on the rank of the random blocks. 
 Our evaluations show a crucial difference between the case of blocks with finite rank $r \geq 1$ and the case of maximal rank $r=d$.\\
It is remarkable that in the most simple formulations of ensembles of sparse random block matrices, with blocks of finite rank, the limiting spectral distribution is 
the Effective Medium Approximation.\\  
The symmetries of the joint probability distribution of the random matrix entries are relevant for the concentration of the probability measure, then it seems proper to study classes of universality.\\
The method of asserting this universality on each contribution of the spectral moments is rather general and we commented its validity on other ensembles where the tree structure of the random graph is known to dominate the limit $N \to \infty$. Further ensembles where loops cannot be neglected may also be addressed with this method.\\
The double limit considered in this note is often studied with the cavity method. Future work to elucidate the common features of the two methods would be relevant. It would be helpful for applications to the very large area of network models.

 \section{Acknowledgment}
	
	 G.M. Cicuta thanks Graziano Vernizzi for discussions on high dimensional spaces.

 \end{document}